1# A Dynamic Watermarking Technique for Matching Communication Addresses with Cars in a Visual Field

Woo-Hyun Ko, Jaewon Kim, Tzu-Hsiang Lin, Samin Moosavi, and P. R. Kumar, *Fellow, IEEE**Abstract*— We consider a problem faced by an intelligent roadside unit (RSU) monitoring a roadway by a video camera. Suppose the RSU notices that a particular car in its visual field needs to execute a specific evasive maneuver to avoid danger. It would like to send a packet addressed to that particular car with this suggestion. The problem is that while all the cars are communicating with the RSU, the RSU does not know which car in the video is associated with what IP address. So, it does not know which IP address to send the packet to. Indeed, the problem of matching addresses with cars in the visual field is a fundamental open problem. We provide an active solution employing "dynamic watermarking" that was originally developed for the security of cyber-physical systems. This technique calls for a car to superpose a small random excitation onto its actuation commands for steering angle or throttle/brake positions. The car sends this random waveform to the RSU in a packet containing its IP address. By signal processing of a car's video stream at the RSU it can verify whether it matches with the waveform in the packet and thereby associates that packet's IP address with that car in the visual field. The RSU thereby determines which IP address is associated with which car in its visual field. We present three demonstrations of performance. We demonstrate experimental results on a laboratory transportation testbed featuring automated vehicles, a vision system, and a network, as well as on the field with two passenger sedans in practice. The results demonstrate that employing the dynamic watermarking method enables an RSU to distinguish a target vehicle's communication from that of other IP addresses of nearby vehicles.

*Index Terms*—Roadside Unit, V2I, Intelligent Vehicle Identification, IP address.## I. INTRODUCTION

INTELLIGENT infrastructure can have various sensory systems to observe vehicles, pedestrians, and the environment on or around the road to obtain situational awareness in real-time (Fig. 1). Once it deduces any potentially dangerous situation, warning signals can be immediately transmitted to the related vehicles. These advanced systems can rely on roadside units (RSUs) [1] to provide traffic information and situational awareness to the vehicles over vehicular ad-hoc networks [2]. While RSUs are usually used to support vehicular ad-hoc networks as a hub or an arbiter [3], one of the important roles of the RSUs is to provide vehicle-to-infrastructure (V2I) safety applications as an information provider delivering reports to the vehicles or giving warning signals to the vehicles approaching an imminent danger. In particular, this feature can be very helpful for driver assists using (V2I) communication [4].

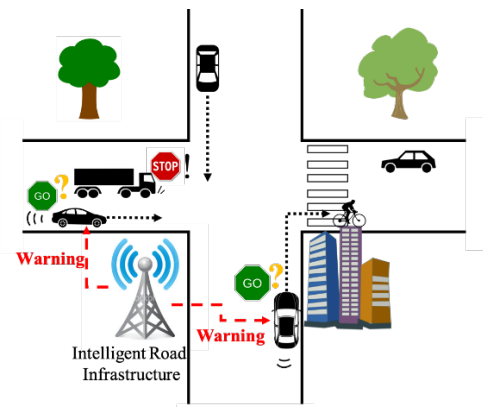

**Fig. 1.** An intelligent road-infrastructure system.

In this paper, we address the problem of how an intelligent RSU can determine the specific communication addresses of the vehicles seen by its visual monitoring system. *For privacy reasons, such addresses may be identifiers in the packet headers of dedicated short-range communication (DSRC) that are used only for a limited temporary period with no permanent assignment [5][6]*. This is a problem of matching information in the cyber layer consisting of the wireless network with visual observation data in the physical layer.

This problem is not reliably solved by each vehicle

This material is based upon work partially supported by the DoD US Army under Contracts W911NF2120064, W911NF2520046, and W911NF2210151, and the DoD Office of Naval Research under Contract N000142412615. The views expressed herein and conclusions contained in this document are those of the authors and should not be interpreted as representing the views or official policies, either expressed or implied, of the U.S. Army Navy, or the United States Government. The U.S. Government is authorized to reproduce and distribute reprints for Government purposes notwithstanding any copyright notation herein.W. Ko is with Texas A&M University System Offices, Bush Combat Development Complex, Bryan, TX 77807 USA (e-mail: whko@tamu.edu). J. Kim is with Texas A&M Global Cyber Research Institute, College Station, TX 77840 USA. (e-mail: j1k@tamu.edu). T. Hsiang is with the Electrical and Computer Engineering Department, Texas A&M University, TX 77845 USA, (e-mail: thl1246@tamu.edu). S. Moosavi is with Texas A&M University System Offices, Bush Combat Development Complex, Bryan, TX 77807 USA (e-mail: saminmoosavi@tamu.edu). P. R. Kumar is with the Electrical and Computer Engineering Department, Texas A&M University, TX 77845 USA, (e-mail: prk@tamu.edu).

broadcasting its GPS coordinates. Such coordinates are often in error [7][8] or have an error margin that, if used in safety situations, can lead to accidents. When vehicles are moving in close proximity, it is not possible to reliably associate the GPS coordinates with the correct vehicle. A vehicle wrongly identified with the incorrect GPS coordinate may be asked to perform a dangerous maneuver, while the right vehicle which needs to perform that maneuver is not informed to perform it. For all these reasons, the "simple" solution of "broadcasting your GPS" is not usable in safety applications.

Another issue is that vehicles may not want to broadcast their position for privacy reasons.

We propose the usage of the technique of "dynamic watermarking" to enable an RSU to find a target vehicle's communication IP address. In prior works, the dynamic watermarking method has been suggested for a different purpose: the security of cyber-physical systems [9][10]. It has been demonstrated how this method can be used to secure vehicles [11], chemical process control systems [13], and the power grid [14][15]. It is helpful to first understand this technique before showing how it can be used to match network and visual data.

We explore the potential of employing this method for a different purpose: enabling an RSU to match individual vehicle position data, observed by its sensory system, to the corresponding communication addresses. To achieve this, we extend the dynamic watermarking technique to the context of multi-vehicle tracking for infrastructure safety systems. In this approach, each vehicle superposes a random private excitation signal on its control inputs and shares this signal with the RSU over a V2I communication network. Upon receiving the private excitation signal, the RSU determines which vehicle's position data is most correlated with the excitation signal by conducting "watermark detection tests" on the position data of all vehicles. Based on the results of these dynamic watermark tests, the RSU identifies the correct communication address for the target vehicle, which the RSU then subsequentially uses to transmit relevant information to the target vehicle.

Through both lab and field experiments, we demonstrate that this technique does effectively enable intelligent infrastructure systems to identify a vehicle's communication address. This enables it to issue warnings to drivers about potential threats, enhancing situational awareness and safety.

Our key contributions are as follows:
- We propose a proactive method using a Dynamic Watermarking technique to match visual IDs of the vehicles observed by a Roadside Unit to their communication addresses used over a V2I network.
- The proposed method requires only implicit data transmission from vehicles, such as the perturbation applied on control inputs, unlike specific position data which can cause a privacy issue. While DSRC may employ such heartbeat messages for other purposes we do not require such privacy violating transmissions.
- We experimentally demonstrate that dynamic watermarking enables the Roadside Unit to successfully match the communication IP addresses to visual IDs of moving vehicles in a comprehensive laboratory scale testbed with a networking system and a vision system.
- We further demonstrate that dynamic watermarking enables the Roadside Unit to successfully match the communication addresses to the visual IDs of moving vehicles in real-world conditions, using full-scale vehicles and a commercial traffic monitoring radar system.

The rest of the paper is organized as follows. In Section II, we review related works on infrastructure safety systems. In Section III, we formulate the problem. After introducing a dynamic watermarking mechanism in Section IV, we describe the design of a real-time vehicle identification algorithm with dynamic watermarking for an intelligent RSU in Section V. Section VI contains the description of the lab experimental setup with overviews of the testbed and middleware and presents an experimental result showing that dynamic watermarking helps to match each vehicle's communication IP address to visually observed odometry information in real-time. The field experimental results with two sedans and a radar system on the runways are shown in Section VII. We conclude in Section VIII.

## II. RELATED WORKS

There has been much research interest in V2I communication technology to solve traffic congestion problems and improving vehicle safety. Early efforts on connected vehicle technology have focused on improving the performance of the Vehicular Ad Hoc Networks (VANETs) by suggesting deployment strategies of RSUs supporting wireless communication among the vehicles [16][17]. Several studies have also focused on the usage of RSU communication with the connected vehicles on the road to increase the intelligence of the system [18-20].

While more recent studies with advanced technologies have proposed the usage of roadside units for intelligent intersection control to enhance road safety features and improve efficiency of transportation systems [21-23], a fundamental problem still remains in their applications: how to figure out the relationship between network-layer information and physical-layer data in the system. That is, how would an RSU determine to which communication address it should transmit pertinent information. For example, if it deduces that a particular vehicle observed by the infrastructure cameras is in danger of an imminent accident, to which particular IP address should it send the information asking it to swerve to avoid the accident?

Numerous studies have focused on the problem of matching network data with visual data. Du et. al have proposed EV-Linker to couple the mobile devices to their owners based on electronic and visual signals by taking advantage of a sniffer and a network camera [24]. They check the consistency of the variances of Received Signal Strength Indicator (RSSI) signals from the mobile devices and observed distances to the owners, caused by the owners' movements. Similarly, Nguyen et. al have proposed IdentityLink to determine links between devices and users by using cameras and RF signals [25]. Two prediction models exploiting a matching likelihood score were also suggested to improve overall linking accuracy between visual and RF patterns. Li et. al have focused on matching human objects in two large visual and electronic datasets for large-scale surveillance [26]. They proposed an EV-Matching algorithm which bridges the two large datasets by splitting Electronic ID sets and finding the corresponding Visual IDs based on spatiotemporal correlation, iteratively and in parallel. One problem with using received signal strength is its unreliability. The received signal depends on multiple scatterers and reflectors in a cluttered environment, and cannot be used in a reliable manner. Moreover, such methods may have greater success with slowly moving human objects rather than with objects like fast moving vehicles.

Lu et. al and Tummala et. al have proposed using a smartphone or dashboard camera deployed in vehicles in a distributed manner [27][28]. They propose an approach to map visual IDs of vehicles observed by the dashboard-mounted camera to communication addresses received through inter-vehicular communication using motion signatures and visual features transmitted. It employs vehicles transmitting information about their colors, models, etc. One of the limitations the proposed RoadMap and ForeSight have in common is that they enable individual vehicles to only have knowledge about the local map of vehicles in their immediate vicinity. To establish a global map, they suggest Roadview in conjunction with these two methods. RoadView has a central server receiving local maps generated by individual vehicles and merging them into the global map to support collaborative vehicular applications related to traffic statistics and safety [29]. This method too relies on the strengths of received radio signals from objects. Moreover, it depends entirely on the explicit data exchanges among vehicles about their real-time positions, colors, models and so on, which raises concerns regarding privacy as well as accuracy.

In this paper, we examine a different approach for multi-vehicle identification. Every vehicle superposes a small random excitation on to its actuators, and the RSU determines which measurements reported by its vision system have the required correlation with the target vehicle's private excitation. Our proposed approach tests two residuals (Test 1 and Test 2) which show that target vehicle can be distinguished from its surrounding vehicles by performing those tests in real-time. We also validate the method in the field.

## III. PROBLEM FORMULAR

With the rapid growth of the number of vehicles, many researchers have developed vehicular technologies for road safety and efficient traffic management. The trend of vehicular technologies mainly concerns two aspects: improving the intelligence of individual vehicles, and sharing information by connecting vehicles. Some studies have proposed situation-awareness approaches using only locally observed data from the mounted sensory systems [30][31]. Other studies have focused on vehicular ad-hoc networks and vehicle-to-infrastructure (V2I) networks to support the communication among vehicles and RSUs to share safety-critical information [32][33].

We address the problem of how an intelligent RSU can send specific information only to a relevant vehicle among vehicles moving on the road over an underlying vehicular ad-hoc network. The RSU communicates with the moving vehicles through the wireless network to provide them with appropriate information, such as road construction, traffic status, accidents ahead, or an urgent situation threatening their safety. Those notifications are crucial for drivers to protect themselves from any dangers unobservable to them. However, to provide a specific service to a particular vehicle in need, the RSU needs to know the communication IP address of that specific vehicle. This is a fundamental problem for intelligent RSUs to match their visual information to communication information.

We formulate that problem as one involving N vehicles on the road and an RSU. We assume that the RSU observes the set consisting of the positions of all the vehicles in its visual field. For warning applications, the RSU needs to warn a particular vehicle about a coming threat by matching its IP address with its position in the RSU's visual field. It is assumed that the network is complete so that an RSU can communicate with every vehicle over the network. Through the underlying communication network, the RSU and individual vehicles exchange data with each other for driver assistance purposes.

In providing roadside assistance, a fundamental problem is that the roadside unit is unable to determine which IP address it should communicate with to alert regarding an imminent situation. It knows the list of all communication IP addresses, but not which IP address belongs to a packet which is about to collide with an obstacle ahead. Broadcasting vehicles' GPS location is not solution. First, the position information can have errors, and it is not safe to rely on it. Also, due to privacy and security issues, vehicles should not be required to broadcast their positions publicly through the shared network. The way to resolve that problem should be in an implicit manner. Since all observed vehicles are connected to the RSU, when the RSU notices the imminent danger for the particular vehicle, the IP address of the vehicle exists in the list of the IP addresses the RSU possesses. Then, the problem becomes how the RSU matches the list of visual observation vehicle IDs to the list of IP addresses.

Our goal in this paper is to effectively assist connected vehicles in securing their safety in a proactive and privacy-safe manner by developing techniques that enable an intelligent road-infrastructure to match the locations of the vehicles obtained by observation data in the physical layer to the IP addresses received through the wireless communication network in the cyber layer.

## IV. DYNAMIC WATERMARKING

In this section, we introduce "Dynamic Watermarking" which is a proactive method to diagnose the purity of the feedback measurement using private excitation signals. Dynamic Watermarking can secure cyber-physical systems against falsification in sensor measurements in [9-12].

In order to explain the concept of dynamic watermarking, we consider a first-order stochastic linear dynamical system which consists of the output, $y \in \mathbb{R}$, the input, $u \in \mathbb{R}$, the known parameters, $a, b \in \mathbb{R}$, and the system noise, $w \sim \mathcal{N}(0, \sigma_w^2)$ which is a Gaussian random variable. The system evolves as:

$$y[t+1] = ay[t] + bu[t] + w[t]. \quad (1)$$

The controller node chooses some policy $g$ and determines a control policy-specified input $u^g[t]$ at every control sampling time $t$. The actuator node generates a random value, $e[t] \sim \mathcal{N}(0, \sigma_e^2)$ called the watermarks whose distribution is i.i.d. of the system output $y[s], 0 \leq s \leq t$. The actual realization of $\{e[t]\}$ is not revealed to any other nodes in the system. That is the reason why the sequence $\{e[t]\}$ is considered as a private excitation applied by the actuator node. The total control input applied by the actuation node to the system is

$$u[t] = u^g[t] + e[t]. \quad (2)$$

By substituting $u[t]$ in (1) with (2), we have

$$y[t+1] - ay[t] - bu^g[t] - be[t] = w[t], \quad (3)$$

as well as

$$y[t+1] - ay[t] - bu^g[t] = be[t] + w[t]. \quad (4)$$

Since the right sides of (3) and (4) are comprised of i.i.d. random variables, the statistics of their left sides, which contain the latest output, $y[t+1]$, are

$$\{y[t+1] - ay[t] - bu^g[t] - be[t]\} \sim \mathcal{N}(0, \sigma_w^2), \quad (5)$$

and

$$\{y[t+1] - ay[t] - bu^g[t]\} \sim \mathcal{N}(0, b^2\sigma_e^2 + \sigma_w^2). \quad (6)$$

The actuator does not have direct access to $y[t]$. Instead there is a sensor that report an output $z[t]$. If $z[t] \equiv y[t]$, the sensor is hones. Otherwise, it is not. To determine which is the case, the controller conducts some test. To check if the reported output $z[t]$ contains the contribution of the private excitation, it conducts the following two tests on the sequence of outputs.

1) Test 1: The actuator node checks if the output of the system satisfies

$$\lim_{T \to \infty} \frac{1}{T} \sum_{k=0}^{T}(z[k] - az[k-1] - bu^g[k-1] - be[k-1])^2$$
$$= \sigma_w^2. \quad (7)$$

2) Test 2: The actuator node checks if the output of the system satisfies

$$\lim_{T \to \infty} \frac{1}{T} \sum_{k=0}^{T}(z[k] - az[k-1] - bu^g[k-1]) = b^2\sigma_e^2 + \sigma_w^2.$$
$$(8)$$

Both these tests are satisfied if the measurements reported by the sensor are correct, i.e., $z[t] \equiv y[t]$. These asymptotic tests can be converted into finite time $x^2$-test in standard ways [12]. The above quantities on the left-hand side are calculated with the reported output $z(\cdot)$ and those values should remain within certain valid ranges depending on $\sigma_w^2$ and $\sigma_e^2$ to pass the two tests. If the sequence of the outputs passes the two watermark tests, it means that the sequence of the outputs is appropriately correlated to the sequence of the corresponding watermarks. This approach provides a systematic method for assessing whether the relationship between measurements and control inputs is appropriate by using a dynamic watermarking mechanism.

V. DESIGN OF REAL-TIME VEHICLE IDENTIFICATION WITH DYNAMIC WATERMARKING FOR INTELLIGENT ROADSIDE UNIT

Based on the above idea of dynamic watermarking, we design a real-time vehicle identification method to enable an intelligent RSU to determine the communication IP addresses of the moving vehicles by using its visual information. We allow an RSU to obtain the position data of the vehicles only from its observation data obtained by its own vision system. We focus on those vehicles that the RSU can both observe and communicate with. We also assume that all the vehicles are honest so that their sharing information is correct.

Fig. 2 shows the design of the dynamic-watermarking based method mapping network IP addresses to visually observed vehicular IDs. Let us suppose there are N moving vehicles on the road. In the case of human-driven vehicles, drivers determine control inputs, whereas, in the case of self-driving vehicles, control inputs are determined by controllers based on the control policy. We consider an actuator, as in Section IV, which superposes a dynamic watermark on the control inputs. At every control sampling time, individual actuators of vehicles superpose a random value onto their own control inputs. Thereby, each vehicle is actuated by the control inputs additively superposed with injected watermarks. The system evolves as:

$$v_i[t+1] = u_{i,v}^g(z_x^t, z_y^t, z_\theta^t) + e_{i,v}[t] + w_{i,v}[t], \quad (9)$$

$$\omega_i[t+1] = u_{i,\omega}^g(z_x^t, z_y^t, z_\theta^t) + e_{i,\omega}[t] + w_{i,\omega}[t], \quad (10)$$

where $v_i[t+1]$ and $\omega_i[t+1]$ are the $i^{\text{th}}$ vehicle's translational velocity and angular velocity, respectively, $u_{i,v}^g(z_x^t, z_y^t, z_\theta^t)$ and $u_{i,\omega}^g(z_x^t, z_y^t, z_\theta^t)$ are control inputs for translational and angular velocities, respectively, $e_{i,v}[t] \sim \mathcal{N}(0, \sigma_{e_{i,v}}^2)$ and $e_{i,\omega}[t] \sim \mathcal{N}(0, \sigma_{e_{i,\omega}}^2)$ are the watermarks superposed on the translational velocity and on the angular velocity control inputs, respectively, $w_{i,v}[t] \sim \mathcal{N}(0, \sigma_{w_{i,v}}^2)$ and $w_{i,\omega}[t] \sim \mathcal{N}(0, \sigma_{w_{i,\omega}}^2)$ are the noises in the translational and angular velocities, respectively, and $z_x^t, z_y^t$ and $z_\theta^t$ are x, y and angular orientation of the vehicle at time t.

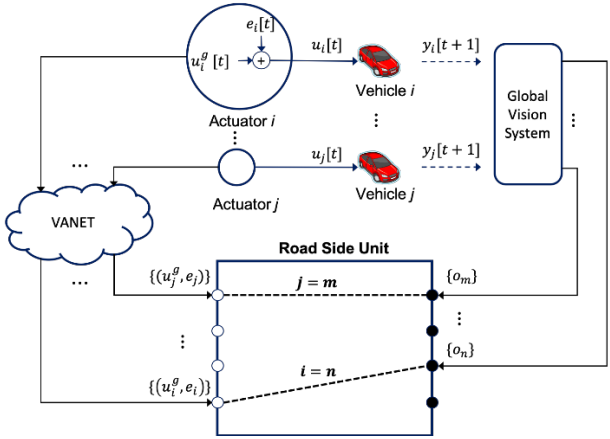

**Fig. 2.** Dynamic-Watermarking based method mapping network IP addresses to visually observed vehicular IDs.

The applied control inputs of all the vehicles are sent to the RSU through the wireless network. The RSU receives a packet including each vehicle's control inputs $u_i^g$ and watermarks $e_i$ through the underlying Vehicular Ad Hoc Network (VANET). Moreover, the RSU receives position data, $o_i \in O = \{(x_i, y_i, \theta_i, t_i) | i = 1, \cdots, N\}$ along with vehicular IDs assigned by the vision system. The RSU then performs two watermark tests on the received information, and the position data of the vehicles obtained by the vision system, to match the communication IP addresses in the network to the corresponding vehicular IDs in the visual data. Since this is a mapping problem, the RSU subjects the watermark tests (7, 8) to each pair of control inputs with watermarks, $\{(u_i^g, e_i)\}$, and observed data $\{o_n\}$ for all pairs. Then, it finds the pair for which the LHSs of (7, 8) are minimum. In the case of the velocity test, we define $V_{1,v}^{i,n}[t] := o_{n,v}[t] - u_{i,v}^g[t-1] - e_{i,v}[t-1]$ and $V_{2,v}^{i,n}[t] := o_{n,v}[t] - u_{i,v}^g[t-1]$. Then, the RSU matches $i^*$ to $n^*$ if they satisfy the following equations:

$$(i_1^*, n_1^*) = \underset{i,n \in \{1,\cdots,N\}}{\arg\min} \lim_{t \to \infty} \frac{1}{t} \sum_{k=0}^{t-1} (V_{1,v}^{i,n}[t])^2, \quad (11)$$

$$(i_2^*, n_2^*) = \underset{i,n \in \{1,\cdots,N\}}{\arg\min} \lim_{t \to \infty} \frac{1}{t} \sum_{k=0}^{t-1} (V_{2,v}^{i,n}[t])^2. \quad (12)$$

If the pair of $i$ and $n$ is not matched, it means that the sequence of the watermarks of the $i^{th}$ vehicle, $\{e_i\}$, is not correlated with the sequence of the $n^{th}$ observed data, $\{o_n\}$. The mismatch results in increasing the values of the LHSs of the watermark tests. Thereby, the RSU can find the mapping relations between the transmitted information from the vehicles over a network and the observed data from the vision system. Thereby, the RSU is enabled to send specific information to the communication IP address of a target vehicle.

## VI. LAB EXPERIMENTAL RESULTS

We evaluate this method in two contexts, first in a laboratory scale setup, and subsequently on the road.

### A. Testbed Setup

We experimentally demonstrate how the dynamic watermarking technique can assist an RSU to match vehicles' communication IP addresses to their visual vehicular IDs in real-time by checking if each vehicle's private excitation signal inputs are correlated to the outputs of its physical system. We set up a testbed as an autonomous multi-vehicle control system in the Cyber-Physical Systems Laboratory at Texas A&M University, shown in Fig. 3. The testbed system consists of a set of miniature vehicles which are remotely controlled by computers as low-level vehicle controllers in Fig. 3. Ten Vicon Boneta 10 cameras are mounted at the top frames to capture the image of the road areas where the miniature vehicles move. The captured images are transmitted to one of the vision processing computers to determine the locations and orientations of the miniature vehicles, and that information is distributed to the low-level vehicle controllers once every 50*ms*. The controllers determine the control inputs for their vehicles to follow desired trajectories, which are specified by a supervisory control layer in the system, a high-level controller.

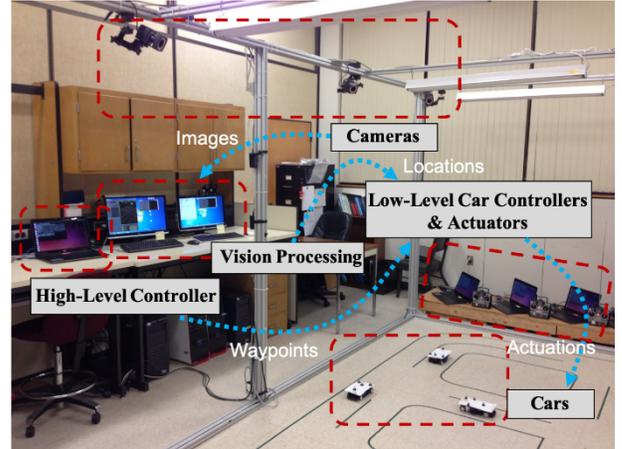

**Fig. 3.** Overview of the laboratory testbed in the Cyber-Physical Systems Lab.

We implement a prototypical intelligent road-infrastructure system of the main interest on the above testbed. The motion of the $i^{th}$ vehicle is described using a kinematic model:

$$x_i[t+1] = x_i[t] + \Delta t \cdot \cos(\theta_i[t]) \cdot (v_i[t] + w_{i,v}[t]), \quad (13)$$

$$y_i[t+1] = y_i[t] + \Delta t \cdot \sin(\theta_i[t]) \cdot (v_i[t] + w_{i,v}[t]), \quad (14)$$

$$\theta_i[t+1] = \theta_i[t] + \Delta t \cdot (\omega_i[t] + w_{i,\omega}[t]), \quad (15)$$

where $x_i[t], y_i[t], \theta_i[t]$ are the x-position, y-position, and the orientation of the $i^{th}$ vehicle at time t, $v_i[t]$, $\omega_i[t]$ are the translational and rotational velocities, respectively, which are the applied control inputs to the $i^{th}$ vehicle at time t. The control noises in the translational and rotational velocities are modeled as $w_{i,v}[t] \sim \mathcal{N}(0, \sigma_{w_{i,v}}^2)$ and $w_{i,\omega}[t] \sim \mathcal{N}(0, \sigma_{w_{i,\omega}}^2)$, respectively. The sampling period of the controller is denoted as $\Delta t$, and the random variables $w_{i,v}[t]$ and $w_{i,\omega}[t]$ denote noises in the translational and rotational velocities. In our testbed, we assume that a miniature vehicle is light enough that a DC motor can accelerate it to the desired translational velocity, which is a control input predetermined by a control law, in a very short time interval. Therefore, using a kinematic model, we take the translational and rotational velocities as the control inputs in the system instead of accelerations, and the errors caused by that general assumption are assumed to be included in the additive noises.

*B. Demonstration of Vehicular Identification in Real-time*

In the laboratory prototype, since we implement an automated vehicular control system on behalf of human drivers, the low-level vehicle controller uses Model Predictive Control (MPC) to determine the control inputs, which are the translational and angular velocities of each vehicle every control sampling time. We consider the scenario where two vehicles follow specifical trajectories. Fig. 4 shows the flow of controls and information among components to control two vehicles and operate an RSU with a vision server in the testbed.

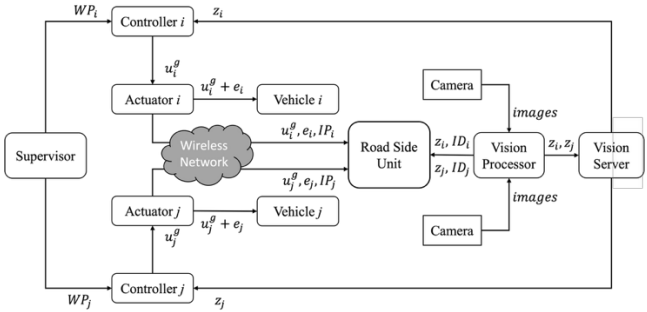

**Fig. 4.** The flow of controls and information among the components in the testbed for real-time vehicular identification.

The Supervisor component schedules and sends trajectory data to two Controller components. The track is an oval path. Vehicles 1 and 2 follow the trajectory. To minimize tracking errors, the MPC Controller components determine control inputs $u_1^g$ and $u_2^g$ for vehicles 1 and 2, respectively. The control inputs specify each vehicle's next speed and the steering angle in the sampled kinematic model. Each control input is sent to the corresponding Actuator component. Employing a Dynamic Watermark, each Actuator component generates a random number with normal distribution as a watermark, $e_i$, and superposes it onto the control inputs. The final actuation commands, $u_1^g + e_1$ and $u_2^g + e_2$ are applied to move vehicle 1 and 2, respectively.

A Vision Sensor component calculates the vehicles' states, $z_1, z_2$, by the images from cameras. A Vision Server component disseminates the information to other components. The Vision Server is implemented in the same computer as the RSU. The RSU receives the position data of vehicles 1 and 2 directly and makes a list of the visual vehicular IDs by assigning $ID_A$ and $ID_B$ to vehicles 1 and 2, respectively and combines the information with the superposed watermark signals obtained over the wireless network to match the vehicles in the visual field with their IP address by using the dynamic watermarking tests.

To discover which visual observation is correlated with which communication information including control inputs and watermarks among two vehicles, we examine the four cases shown in Table 1. The visual observation includes location data of vehicles, while the communication information includes control inputs and watermarks.

TABLE I
VISUAL OBSERVATIONS AND COMMUNICATION INFORMATION USED FOR EACH CASE OF WATERMARK TESTS

| Data Type | Cases | | | |
|---|---|---|---|---|
| | 1 | 2 | 3 | 4 |
| Visual Observation (Location) | $ID_A, z_A$ | $ID_A, z_A$ | $ID_B, z_B$ | $ID_B, z_B$ |
| Communication Info. (Controls, Watermarks) | $IP_1, u_1^g, e_1$ | $IP_2, u_2^g, e_2$ | $IP_1, u_1^g, e_1$ | $IP_2, u_2^g, e_2$ |

Fig. 5 shows the graphs of the experimental results. The variances of watermark 1 and 2 are set to 0.07 and 0.38, respectively. Fig. 5 (a) compares the results of watermark test 1, and Fig. 5 (b) compares test 2 for cases 1 and 3. Case 1 using the correct pair of location data and control inputs with watermarks related to vehicle 1 can be seen to have lower values than case 3 in both tests. Likewise, Fig. 5 (c) compares the results of watermark test 1, and (d) compares the ones of watermark test 2 for cases 2 and 4. They also show that the values of the correct case 4 are lower than the ones of case 2. The RSU can therefore successfully match visual $ID_A$ to communication $IP_1$, and visual $ID_B$ to communication $IP_2$.

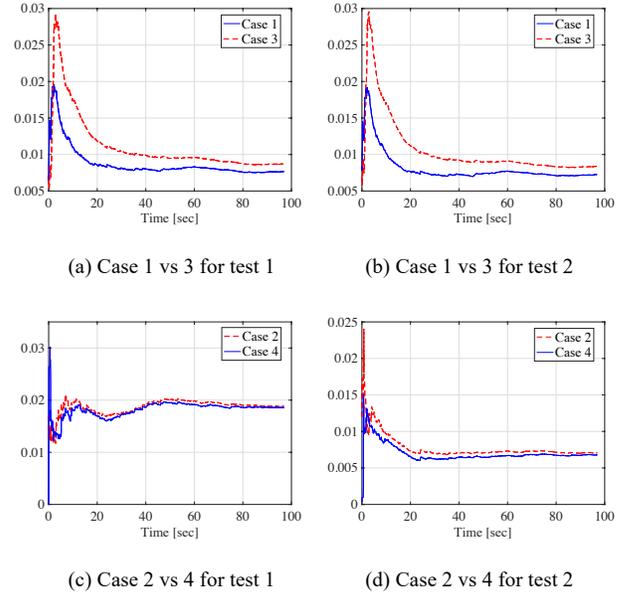

(a) Case 1 vs 3 for test 1    (b) Case 1 vs 3 for test 2

(c) Case 2 vs 4 for test 1    (d) Case 2 vs 4 for test 2

**Fig. 5.** Comparison of the statistical results of the watermark tests for the identification of two vehicles.

The demonstration video clip in [34] shows that by using a dynamic watermarking technique, the RSU can correctly match the control inputs transmitted from each vehicle's communication IP address to the observation data of its vehicular ID from the vision system.

VII. FIELD EXPERIMENTAL RESULTS

To validate the practical employment of our proposed approach, we conducted a field road experiment using real vehicles and a commercial traffic monitoring radar system. The experimental setup, shown in Fig. 7 took place on the runways at RELLIS campus of Texas A&M University. Two cars were used to match their visual IDs, assigned by the radar system, with their communication IDs, IP addresses, that were used to transmit watermarks to the RSU. The self-driving car, a Lincoln MKZ, was equipped with a drive-by-wire system controlled by using a ROS2 trajectory tracking package developed in the lab. The self-driving car followed a predetermined lane while maintaining its target speed by adjusting acceleration and brake pedal positions. Randomized watermarks were injected into the brake values, influencing the self-driving car's velocity. A human-driven car followed the self-driving car at a distance of approximately 2 meters in the same lane.

The RSU system consisted of a radar system and a laptop. The radar system, a Smart Micro Radar Type 48 Stream [35], monitored traffic at the intersection, assigning a unique visual ID to each moving car within its field of view. It also tracked the location and velocity of each car along the lane, updating every 100 milli-seconds. The laptop processed the velocity data from both vehicles, applying dynamic watermark tests to determine which sequence was associated with the watermarks injected into the self-driving car's brake values, as the identity of self-driving car was initially unknown.

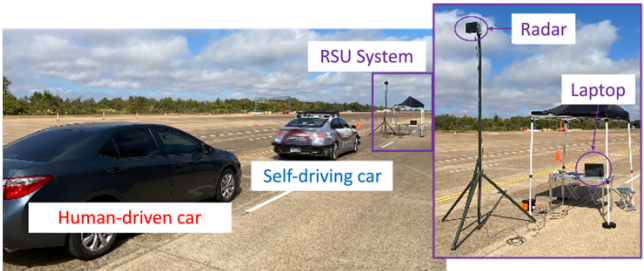

**Fig. 7.** Setup for a field experiment for matching visual IDs from a radar system to communication IDs of a self-driving car and a human-driven car.

The top graph in Fig. 8 illustrates the speed variations of both self-driving and human-driven cars. The x-axis represents time in seconds, while the y-axis shows speed in meters per second.

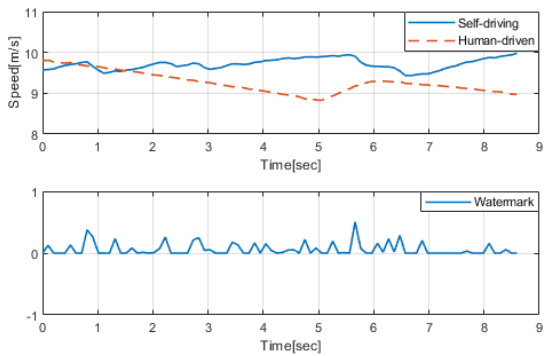

**Fig. 8.** The variations of the speeds of self-driving and human-driven cars (a) and the watermarks imposed to the brake inputs of the self-driving car.

The blue line indicates the speed of the self-driving car, which was set to a target speed of 10 m/s, and the red dotted line represents the speed of the human-driven car, which followed the self-driving car at a constant distance. The bottom graph displays the watermark values, which were randomly generated with 0.05 variance every 100 milli-seconds and applied to the brake inputs of the self-driving car.

We subjected the speed measurements of both the self-driving and the human-driven cars, as recorded by the radar system of the RSU, to dynamic watermark tests. In Fig. 9, Case 1 represented by the blue line show the test results for the self-driving car's speed measurements, while Case 2 represented by the red dotted line corresponds to the human-driven car's speed measurements. Their test values were averaged with a moving window of 20. The more the measurements are associated with the watermark values, the lower the test results should be. This means that the speed measurements of the self-driving car in Case 1, which show lower values than those in Case 2, are more strongly associated with the injected watermarks.

These field experimental results demonstrate that our proposed Dynamic Watermarking technique successfully matched the visual ID of the self-driving car, as assigned by the radar system, to the communication ID of the self-driving car, which transmitted the watermarks correctly.

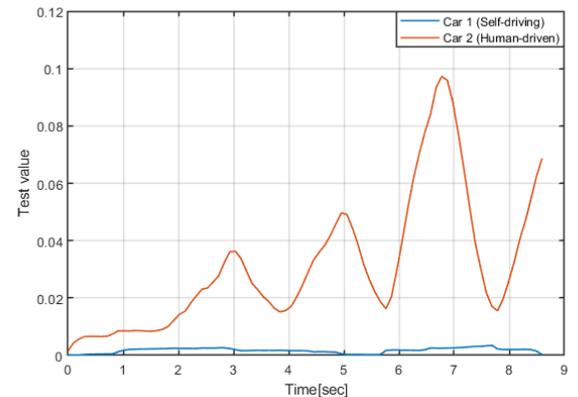

**Fig. 9.** The results of the watermark tests with observed speeds of the self-driving and the human-driven cars.

## VIII. CONCLUSION

This paper addresses the problem of securing safety and sharing data among connected vehicles. We consider an intelligent road-infrastructure system which communicates wirelessly with moving vehicles on the road. RSUs are deployed not only to support communication among the vehicles over the VANET by collecting and spreading useful information and warning signals around them, but also to monitor road situations by using sensory systems. If an RSU notices an imminent situation, it needs to immediately send a warning signal to the relevant vehicle. We employ a dynamic watermarking technique in the RSU to find the communication IP addresses of the vehicles, to which the RSU then needs to transmit packets containing information to enhance road safety. We design a dynamic-watermarking based approach which successfully maps communication IP addresses in the network to vehicular IDs in the visual field. We experimentally demonstrate the proposed dynamic-watermarking based method's ability to find the communication IP addresses for two automated vehicles in a laboratory testbed. We also present field experimental results showing that the proposed method is successful with real-world vehicles and a traffic monitoring radar system.

REFERENCES

[1] U.S. Department of Transportation, "An Overview of USDOT Roadside Unit Research," [Online]. Available: https://rosap.ntl.bts.gov/view/dot/34763

[2] NXP, "Intelligent Roadside Unit," [Online]. Available: https://www.nxp.com/applications/solutions/automotive/connectivity/intelligent-roadside-unit:INTELLIGENTRSU

[3] Siemense, "Connected Vehicle Roadside Unit," [Online]. Available: https://w3.usa.siemens.com/mobility/us/en/road-solutions/traffic-management/Documents/Siemens%20RSU%20Brochure_NEW.pdf

[4] V. A. Butakov and P. Ioannou, "Personalized Driver Assistance for Signalized Intersections Using V2I Communication", IEEE Transactions on Intelligent Transportation Systems, Vol. 17, Issue 7, July 2016, pp. 1910 - 1919.

[5] J. Esposito, "An Application of DSRC Between Connected Vehicles and Intelligent Transportation Systems," Thesis for the degree of master of science, University of Florida.

[6] C. Bettisworth, M. Burt, A. Chachich, R. Harrington, J. Hassol, A. Kim, K. Lamoureux, D. LaFrance-Linden, C. Maloney, D. Perlman, G. Ritter, S. M. Sloan, and Eric Wallischeck, "Status of the Dedicated Short-Range Communications Technology and Applications", U.S. Department of Transportation Technical report to Congress, July 2015.

[7] P. Ranacher, R. Brunauer, W. Trutschnig, S. V. Spek and S. Reich, "Why GPS makes distances bigger than they are", International Journal of Geographical Information Science, 2016, Vol. 30, No. 2, pp. 316–333.

[8] B. A. Renfro, M. Stein, N. Boeker and A. Terry, "An Analysis of Global Positioning System (GPS) Standard Positioning Service (SPS) Performance for 2017", Space and Geophysics Laboratory Applied Research Laboratories, the University of Texas at Austin, March 20, 2018.

[9] S. Weerakkody, Y. Mo, and B. Sinopoli, "Detecting integrity attacks on control systems using robust physical watermarking," in 53rd IEEE Conference on Decision and Control, Dec 2014, pp. 3757–3764.

[10] B. Satchidanandan and P. R. Kumar, "Secure control of networked cyber-physical systems." 2016 IEEE 55th Conference on Decision and Control (CDC), pp. 283–289, Dec 2016.

[11] B. Satchidanandan and P. R. Kumar, "Dynamic watermarking: Active defense of networked cyber-physical systems." Proceedings of the IEEE 105, 2, Feb 2017, pp. 219–240.

[12] W. Ko, B. Satchidanandan and P. R. Kumar, "Theory and implementation of dynamic watermarking for cybersecurity of advanced transportation systems." 2016 IEEE Conference on Communications and Network Security (CNS), pp. 416–420, Oct 2016.

[13] J. Kim, W. Ko and P. R. Kumar, "Cyber-security with Dynamic Watermarking for Process Control Systems", AIChE Annual Meeting, Orlando, FL, Nov. 10-15, 2019.

[14] T. Huang, B. Satchidanandan, P. R. Kumar and L. Xie, "An Online Detection Framework for Cyber Attacks on Automatic Generation Control," IEEE Transactions on Power Systems, pp. 6816-6827, Vol. 3, Issue 6, November 2018.

[15] W. Ko, J. A. Ramos-Ruiz, T. Huang, J. Kim, H. Ibrahim, P. N. Enjeti, P. R. Kumar, and L. Xie, "Robust Dynamic Watermarking for Cyber-Physical Security of Inverter-based Resources in Power Distribution Systems", IEEE Transactions on Industrial Electronics, Vol. 71, Issue 7, pp. 7106-7116, July 2024

[16] J. Barrachina, P. Garrido, M. Fogue, F. J. Martinez, J.-C. Cano, C. T. Calafate and P. Manzoni, "Road Side Unit Deployment: A Density-Based Approach", IEEE Intelligent Transportation Systems Magazine, Vol.5, Issue 3, pp. 30-39, 22 July 2013.

[17] M. Fogue, J. A. Sanguesa, F. J. Martinez and J. M. Marquez-Barja, "Improving Roadside Unit Deployment in Vehicular Networks by Exploiting Genetic Algorithms", Applied Sci., 2018, Vol. 8, Issue1-86.

[18] A. Al-Dweik, R. Muresan, M. Mayhew and M. Lieberman, "IoT-based multifunctional Scalable real-time Enhanced Road Side Unit for Intelligent Transportation Systems", 2017 IEEE 30th CCECE, Windsor, ON, Canada, 30 April-3 May 2017.

[19] G. S. Khekare and A. V. Sakhare, "A smart city framework for intelligent traffic system using VANET", 2013 iMac4s, Kottayam, India, 22-23 March 2013.

[20] L.Sun, Y. Wu, J. Xu and Y. Xu, "An RSU-assisted localization method in non-GPS highway traffic with dead reckoning and V2R communications", 2012 2nd International Conference on CECNet, pp. 149-152, Yichang, China, 21-23 April 2012.

[21] M. Evans, M. Machado, R. Johnson, A. Vadella, L. Escamilla, B. Froemming-Aldanondo, T. Rastoskueva, M. Jostes, D. Butani, R. Kaddis, C. Chung, and J. Siegel, "Vehicle-to-Everything (V2X) Communication: A Roadside Unit for Adaptive Intersection Control of Autonomous Electric Vehicles", arXiv:2409.00866 [cond-mat] (2024) (available at https://arxiv.org/abs/2409.00866)

[22] K. Wang, C. She, Z. Li, T. Yu, Y. Li, and K. Sakaguchi, "Roadside Units Assisted Localized Automated Vehicle Maneuvering: An Offline Reinforcement Learning Approach", arxiv:2405.03935 [cond-mat] (2024) (available at https://arxiv.org/abs/2405.03935)

[23] P. Sun, X. Qi, and R. Zhong, "A Roadside Precision Monocular Measurement Techonology for Vehicle-to-Everything (V2X)", Sensors 2024, 24(17), 5730.

[24] S. Du, J. Hua, Y. Gao and S. Zhong, "EV-Linker: Mapping eavesdropped Wi-Fi packets to individuals via electronic and visual signal matching", Journal of Computer and System Sciences 82 (2016), pp. 156-172.

[25] L. T. Nguyen, Y. S. Kim, P. Tague and J. Zhang, "IdentityLink: User-Device Linking through Visual and RF-Signal Cues", UBICOMP '14, Sep. 13-17, 2014, Seattle, WA, USA.

[26] G. Li, F. Yang, G. Chen, Q. Zhai, X. Li, J. Teng, J. Zhu, D. Xuan, B. Chen and W. Zhao, "EV-Matching: Bridging Large Visual Data and Electronic Data for Efficient Surveillance", IEEE 37th ICDCS, pp. 689-698, Atlanta, GA, USA, June 5-8, 2017.

[27] D. Li, Z. Lu, T. Bansal, E. Schilling and P. Sinha, "ForeSight: Mapping vehicles in visual domain and electronic domain", IEEE INFOCOM 2014, Toronto, ON, Canada, 27 April-2 May 2014.

[28] G. K. Tummala, D. Li and P. Sinha, "RoadMap: mapping vehicles to IP addresses using motion signatures", ACM CarSys 2016, pp. 30-37, 3 Oct. 2016.

[29] G. K. Tummala, D. Li and P. Sinha, " Roadview: Live View of On-Road Vehicular Information", IEEE SECON 2017, San Diego, CA, USA, 12-14 June 2017.

[30] D. Dong, X. Li and X. Sun, "A Vision-Based Method for Improving the Safety of Self-Driving", 12th International Conference on Reliability, Maintainability, and Safety (ICRMS), Shanghai, China, 17-19 Oct. 2018.

[31] Y. Zhang, J. Wang, X. Wang and J. M. Dolan , "Road-Segmentation-Based Curb Detection Method for Self-Driving via a 3D-LiDAR Sensor", IEEE Transactions on Intelligent Transportation Systems, Vol. 19, Issue 12, Dec. 2018, pp. 3981 - 3991.

[32] R. Coppola and M. Morisio, "Connected Car: Technologies, Issues, Future Trends", ACM Computing Surveys (CSUR), Vol. 49, Issue 3, No. 46, Dec. 2016.

[33] D.-F. Xie, X.-M. Zhao and Z. He, "Heterogeneous Traffic Mixing Regular and Connected Vehicles: Modeling and Stabilization", IEEE Trans. on Intelligent Transportation Systems, Vol. 20, Issue 6, June 2019.

[34] https://youtu.be/pfiPnYOgP1o?si=6oms8t0jXiW67ghX [Online].

[35] Smart Micro Traffic Management Sensor: TRUGRD UMRR-12 Type 48, https://www.smartmicro.com/fileadmin/media/Downloads/Traffic_Radar/Sensor_Data_Sheets__24_GHz_/TRUGRD_UMRR-12_Type_48_Traffic_Management_Datasheet.pdf

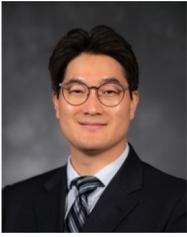
**Woo-Hyun Ko** received the B.S. degree in electrical electronics engineering from Yonsei University, Seoul, Korea in 2005 and the M.S. degree in electrical and computer engineering from Seoul National University, Seoul Korea in 2007. He earned his Ph.D. in electrical engineering at Texas A&M University in 2017. He is currently a Senior Research Engineer at Bush Combat Development Complex. His research interests include cyber-physical systems and cybersecurity, middleware, wireless networks, 5G, power systems, multi-agent systems, and machine learning.

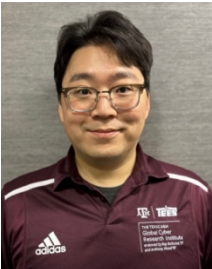
**Jaewon Kim** (Member, IEEE) received the Ph.D. degree in Computer Engineering from Texas A&M University, College Station, TX, USA, in 2023.

He is currently a Research Scientist at Texas A&M Global Cyber Research Institute (GCRI). Before joining GCRI, he was a Postdoctoral Associate in the Laboratory for Information & Decision Systems (LIDS) at Massachusetts Institute of Technology (MIT). His research focus includes cyber-physical systems, cyber-security, network security, machine learning, reinforcement learning, system identification, multi-agent unmanned vehicle systems, resilient real-time network architectures for unmanned aerial/ground vehicles, cloud/fog robotics, and network authentication/segmentation.

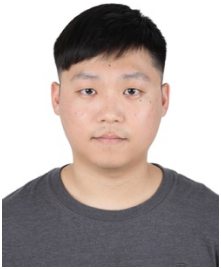
**Tzu-Hsiang Lin** received a bachelor's degree from National Chung Hsing University, Taichung, Taiwan, in 2018. He is currently pursuing a Ph.D. at Texas A&M University, with research interests in cybersecurity, cyber-physical systems, and machine learning.

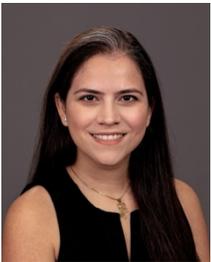
**Samin Moosavi** earned her bachelor's degree in Mechanical Engineering from Sharif University of Technology, Tehran, Iran. She received her Ph.D. degree in Mechanical Engineering at Texas A&M University in 2024.

Her research interests are in sensor fusion, localization, and tracking of autonomous cars. Currently, she is a senior research engineer at the Connected Autonomous Safe Transportation lab, where she is working on different projects related to resilient sensor fusion in autonomous vehicle localization.

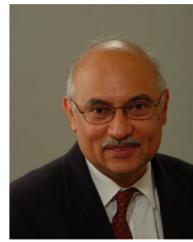
**P. R. Kumar** B. Tech. (IIT Madras, '73), D.Sc. (Washington University, St. Louis, '77), was a faculty member at UMBC (1977-84) and Univ. of Illinois, Urbana-Champaign (1985-2011). He is currently at Texas A&M University. His current research is focused on machine learning, cybersecurity, power systems, 5G, unmanned air vehicle systems, cyberphysical systems and wireless networks.

He is a member of the World Academy of Sciences, the US National Academy of Engineering, and the Indian National Academy of Engineering. He was awarded a Doctor Honoris Causa by ETH, Zurich. He has received the IEEE Alexander Graham Bell Medal, the IEEE Field Award for Control Systems, the Donald P. Eckman Award of the American Automatic Control Council, the Fred W. Ellersick Prize of the IEEE Communications Society, the Outstanding Contribution Award of ACM SIGMOBILE, the Infocom Achievement Award, the SIGMOBILE Test-of-Time Paper Award, and the COMSNETS Outstanding Contribution Award. He is a Fellow of IEEE and an ACM Fellow. He was Leader of the Guest Chair Professor Group on Wireless Communication and Networking at Tsinghua University, was a D. J. Gandhi Distinguished Visiting Professor at IIT Bombay, and is an Honorary Professor at IIT Hyderabad. He was awarded the Distinguished Alumnus Award from IIT Madras, the Alumni Achievement Award from Washington Univ., and the Daniel Drucker Eminent Faculty Award from the College of Engineering at the Univ. of Illinois.